\documentclass[aps,prl,twocolumn,groupedaddress]{revtex4}
\usepackage{graphicx}

\newcommand{\bra}{\langle}
\newcommand{\ket}{\rangle}

\begin{document}


\title{A linear response approach to the calculation of the effective
interaction parameters in the LDA+U method}


\author{Matteo Cococcioni\footnote{present address: Massachusetts Institute 
of Technology, 77 Massachusetts avenue, Cambridge MA, 02139 USA.}
and Stefano de Gironcoli}
\affiliation{SISSA -- Scuola Internazionale Superiore di Studi Avanzati and \\
INFM-DEMOCRITOS National Simulation Center,\\
via Beirut 2-4, I-34014 Trieste, Italy}


\date{\today}

\begin{abstract}
In this work we reexamine the LDA+U method of Anisimov and coworkers in
the framework of a plane-wave pseudopotential approach. A simplified
rotational-invariant formulation is adopted.  The calculation of the
Hubbard $U$ entering the expression of the functional is discussed and
a linear response approach is proposed that is internally consistent
with the chosen definition for the occupation matrix of the relevant
localized orbitals.
In this way we obtain a scheme whose functionality should not depend
strongly on the particular implementation of the model in ab-initio
calculations.
We demonstrate the accuracy of the method, computing structural and
electronic properties of a few systems including transition and
rare-earth correlated metals, transition metal monoxides and iron-silicate.
\end{abstract}


\maketitle

\section{Introduction}

The description and understanding of electronic properties of strongly
correlated materials is a very important and long standing problem for
ab-initio calculations. Widely used approximations for the exchange
and correlation energy in density functional theory (DFT), mainly
based on parametrization of (nearly) homogeneous electron gas, miss
important features of their physical behavior.  For instance both
local spin-density approximation (LSDA) and spin-polarized generalized
gradient approximations ($\sigma-$GGA), in their several flavors, fail in
predicting the insulating behavior of many simple transition metal oxides
(TMO), not only by severely underestimating their electronic band gap but,
in most cases, producing a qualitatively wrong metallic ground state.

TMOs have represented for long time the most notable failure of DFT.
When the high-T$_{{\rm c}}$ superconductors entered the scene (their
parent materials are also strongly correlated systems) the quest for
new approaches that could describe accurately these systems by first
principles received new impulse, and in the last fifteen years many
methods were proposed in this direction.  Among these, LDA+U approach,
first introduced by Anisimov and coworkers \cite{AZA91,ani2,SDA94},
has allowed to study a large variety of strongly correlated compounds
with considerable improvement with respect to LSDA or $\sigma-$GGA
results. The successes of the method have led to further developments
during the last decade which have produced very sophisticated theoretical
approaches\cite{DMFT} and efficient numerical techniques.

The formal expression of LDA+U energy functional is adapted from model
hamiltonians (Hubbard model in particular) that represent the "natural"
theoretical framework to deal with strongly correlated materials. As in
these models, a small number of localized orbitals is selected and the
electronic correlation associated to them is treated in a special way. The
obtained results strongly depend on the definition of the localized
orbitals and on the choice of the interaction parameters used in the
calculation, that should be determined in an internally consistent
with. This is not always done and a widespread but, in our opinion,
unsatisfactory attitude is to determine the value of the electronic
couplings by seeking a good agreement of the calculated properties with
the experimental results in a semiempirical way.

In this work a critical reexamination of the LDA+U approach is
proposed, which starting from the formulation of Anisimov and
coworkers \cite{AZA91,ani2,SDA94}, and its further improvements
\cite{LAZ95,ani5,Pickett98}, develops a simpler approximation. This is,
in our opinion, the "minimal" extension of the usual approximate DFT
(LDA or GGA) schemes needed when atomic-like features are persistent in
the solid environment.

In the central part of this work we describe a method, based on a linear
response approach, to calculate in an internally consistent way---without
aprioristic assumption about screening and/or basis set employed in the
calculation---the interaction parameters entering the LDA+U functional
used. In this context our plane-wave pseudopotential (PWPP) implementation
of the LDA+U approach is presented and discussed in some details.
We stress however that the proposed method is basis-set independent.

Our methodology is then applied to the study of the electronic properties
of some real materials, chosen as representative of "normal" (bulk iron)
and correlated (bulk cerium) metals, as well as a few examples of strongly
correlated systems (iron oxide, nickel oxide and fayalite).

\section{Standard LDA+U implementation:}

In order to account explicitly for the on-site Coulomb interaction
responsible for the correlation gap in Mott insulators and not treated
faithfully within LDA, Anisimov and coworkers \cite{AZA91,ani2,SDA94} correct
the standard functional adding an on-site Hubbard-like 
interaction, $E_{Hub}$:
\begin{eqnarray}
\label{simpleu}
E_{LDA+U}[n({\bf r})] &=& E_{LDA}[n({\bf r})] + \nonumber \\
&&E_{Hub}[\{n^{I\sigma}_{m}\}] -
E_{dc}[\{n^{I\sigma}\}]
\end{eqnarray}
where $n({\bf r})$ is the electronic density, and $n^{I\sigma}_{m}$
are the atomic-orbital occupations for the atom $I$ experiencing
the "Hubbard" term. The last term in the above equation is then
subtracted in order to avoid double counting of the interactions 
contained both in $E_{Hub}$ and, in some average way, in $E_{LDA}$.
In this term the total, spin-projected, occupation of the localized
manifold is used: $n^{I\sigma}=\sum_{m}n^{I\sigma}_{m}$.

In its original definition the functional defined in Eq. \ref{simpleu}
was not invariant under rotation of the atomic-orbital basis set used
to define the occupancies $n^{I\sigma}_{m}$. A rotationally invariant
formulation has then been introduced \cite{LAZ95,ani5} where the orbital
dependence of $E_{Hub}$ is borrowed from atomic Hartree-Fock with renormalized 
slater integrals:
\begin{eqnarray}
\label{ub1}
E_{Hub}[\{n^{I}_{mm'}\}] &=& 
                   \frac{1}{2}\sum_{\{m\},\sigma,I}\{ \bra m,m''|V_{ee}|m',m'''\ket
                               n^{I\sigma}_{mm'}n^{I-\sigma}_{m''m'''}  \nonumber \\
&&+ (\bra m,m''|V_{ee}|m',m'''\ket \nonumber \\
&&-  \bra m,m''|V_{ee}|m''',m'\ket) n^{I\sigma}_{mm'}n^{I\sigma}_{m''m'''} \}
\end{eqnarray}
with
\[
\bra m,m''|V_{ee}|m',m'''\ket = \sum_{k=0}^{2l} a_{k}(m,m',m'',m''') F^{k}
\]
where $l$ is the angular moment of the localized ($d$ or $f$) electrons and
\[
a_{k}(m,m',m'',m''') = \frac{4\pi}{2k+1}\sum_{q=-k}^{k}
\bra lm|Y_{kq}|lm'\ket \bra lm''|Y^{*}_{kq}|lm'''\ket .
\]

The double-counting term $E_{dc}$ is given by:
\begin{eqnarray}
\label{ub2}
E_{dc}[\{n^{I}\}] &=&\sum_{I} \frac{U}{2}n^{I}(n^{I}-1) \nonumber \\
&-&\sum_{I}\frac{J}{2}
[n^{I\uparrow}(n^{I\uparrow}-1)+n^{I\downarrow} (n^{I\downarrow}-1)] .
\end{eqnarray}
The radial Slater integrals $F^{k}$ are the parameters of the model
($F^0$,$F^2$ and $F^4$ for $d$ electrons, while also $F^6$ must be
specified for $f$ states) and are usually re-expressed in terms of 
only two parameters, $U$ and $J$, describing screened on-site Coulomb 
and exchange interaction,
\begin{eqnarray}
\label{ueff}
U &=& \frac{1}{(2l+1)^{2}} \sum_{m,m'}
\bra m,m'|Vee|m,m'\ket = F^{0} \\
J &=& \frac{1}{2l(2l+1)} \sum_{m \ne m',m'}
\bra m,m'|Vee|m',m\ket \nonumber = \frac{F^{2}+F^{4}}{14}, \nonumber
\end{eqnarray}
by assuming atomic values for $F^{4}/F^{2}$ and $F^{6}/F^{4}$ ratios.

To obtain $U$ and $J$, Anisimov and coworkers \cite{AnGun91,SDA94} propose
to perform LMTO calculations in supercells in which the occupation of the
localized orbitals of one atom is constrained. The localized orbitals
of all atoms in the supercell are decoupled from the remainder of the
basis set. This makes the treatment of the local orbitals an atomic-like
problem---making it easy to fix their occupation numbers---and allows to
use Janak theorem \cite{Janak} to identify the shift in the corresponding
eigenvalue with the second-order derivative of the LDA total energy with
respect to orbital occupation. It has however the effect of leaving
a rather artificial system to perform the screening, in particular
when it is not completely intra-atomic. In elemental metallic Iron,
for instance, Anisimov and Gunnarsson \cite{AnGun91} showed that
only half of the screening charge is contained in the Wigner-Seitz
cell. This fact, in addition to a sizable error due to the Atomic
Sphere Approximation used \cite{AnGun91}, could be at the origin of the
severe overestimation of the computed on-site coulomb interaction with
respect to estimates based on comparison of spectroscopic data and model
calculations\cite{ExpFeU1,ExpFeU2}.

\section{Basis set independent formulation of LDA+U method}

Some aspects of currently used LDA+U formulation, and in particular
of the determination of the parameters entering the model, have been so far
tied to the LMTO approach.
This is not a very pleasant situation and some efforts have been done
recently \cite{Pickett98,Bengone00} to reformulate the method for
different basis sets.
Here we want to elaborate further on these attempts and provide an
internally consistent, basis-set independent, method for the calculation
of the needed parameters.

\subsection{Localized orbital occupations}

In order to fully define how the approach works the first thing to do is
to select the degrees of freedom on which ``Hubbard $U$" will operate
and define the corresponding occupation matrix, $n^{I\sigma}_{mm'}$.
Although it is usually straightforward to identify in a given system the 
atomic levels to be treated in a special way (the $d$ electrons in 
transition metals and the $f$ ones in the rare earths and actinides series) 
there is no
unique or rigorous way to define occupation of localized atomic levels in a 
multi-atom system.
Equally legitimate choices for $n^{I\sigma}_{mm'}$ are $i)$ projections 
on normalized atomic
orbitals, or $ii)$ projections on Wannier functions whenever
the relevant orbitals give
raise to isolated band manifolds, or $iii)$ Mulliken population or $iv)$
integrated values in (spherical) regions around the atoms of the
angular-momentum-decomposed charge densities.
Taking into account the arbitrariness in the definition of
$n^{I\sigma}_{mm'}$ no particular significance should be attached to
any of them (or other that could be introduced) and the usefulness and
reliability of an approximate DFT+U method (aDFT+U), and of its more
recent and involved evolutions like the aDFT+DMFT method, should be
judged from its ability to provide a correct physical picture of the
systems under study irrespective of the details of the formulation,
once all ingredients entering the calculation are determined consistently.

All above mentioned definitions for the occupation matrices can be put in the 
generic form
\begin{equation}
\label{eq:nijdef}
n^{I\sigma}_{mm'} = \sum_{{\rm \bf k},v}f^\sigma_{{\rm \bf k}v}
\bra \psi^{\sigma}_{{\rm \bf k}v}| P^I_{mm'}|\psi^{\sigma}_{{\rm \bf k}v}\ket
\end{equation}
where $\psi^{\sigma}_{{\rm \bf k}v}$ is the valence electronic
wavefunction corresponding to the state (${\rm \bf k}v$) with spin
$\sigma$ of the system and $f^\sigma_{{\rm \bf k}v}$ is the corresponding 
occupation number. 
The $P^I_{mm'}$'s are generalized projection operators on the
localized-electron manifold that satisfy the following properties:
\begin{eqnarray}
\nonumber & \sum_{m'} P^I_{mm'} P^I_{m'm''} = P^I_{mm''}; 
\quad P^I_{mm'} = (P^I_{m'm})^\dagger ; &\\
& P^I_{mm'}P^I_{m''m'''} = 0 \quad {\rm when} \quad m'\not=m'' . &
\end{eqnarray}
In particular $P^I = \sum_m P^I_{mm} $ is the projector on the complete
manifold of localized states associated with atom at site $I$ and
therefore
\begin{equation}
\label{eq:nI}
n_I = \sum_{\sigma} \sum_{{\rm \bf k},v}f^\sigma_{{\rm \bf k}v}
\bra \psi^{\sigma}_{{\rm \bf k}v}| P^I|\psi^{\sigma}_{{\rm \bf k}v}\ket
=\sum_{\sigma,m} n^{I\sigma}_{mm}
\end{equation}
is the total localized-states occupation for site $I$.
Orthogonality of projectors on different sites is {\it not} assumed.

In the applications discussed in this work we will define localized-level
occupation matrices projecting on atomic pseudo-wavefunctions.
The needed projector operators are therefore simply
\begin{equation}
P^I_{mm'} = |\varphi^I_m\ket \bra \varphi^I_{m'}|
\end{equation}
where $|\varphi^{I}_m\ket$ is the valence atomic orbital with angular
momentum component $|lm\ket$ of the atom sitting at site $I$ (the
same wavefunctions are used for both spins).  Since we will be using ultrasoft
pseudopotentials to describe valence-core interaction, all scalar
products between crystal and atomic pseudo-wavefunctions are intended
to include the usual S matrix describing orthogonality in presence of
charge augmentation \cite{USPP}.

As already mentioned, other choices could be used as well and different
definitions for the occupation matrices will require, in general,
different values of the parameter entering the aDFT+U functional, as it
has been pointed out recently also by Pickett et al. \cite{Pickett98}
where, for instance, the value of Hubbard $U$ in FeO shifts from 4.6 to
7.8 eV when atomic d-orbitals for Fe$^{2+}$ ionic configuration are used
instead of those of the neutral atom. In an early study \cite{McMahan88}
the U parameter in La$_2$CuO$_4$ varies from 6.8 to 7.7 eV upon variation
of the atomic sphere radius employed in the LMTO calculation. As pointed
out in these works it is not fruitful to compare numerical values of
U obtained by different methods but rather comparison should be made
between results of complete calculations.

\subsection{A simplified rotationally invariant scheme and the meaning of U}

In order to simplify our analysis and gaining a more transparent
physical interpretation of the "+U" correction to standard aDFT
functionals we concentrate on the main effect associated to on-site
Coulomb repulsion. We thus neglect the important but somehow secondary effects
associated to non sphericity of the electronic interaction and the proper
treatment of magnetic interaction, that in the currently used rotational
invariant method is dealt with assuming a screened Hartree-Fock form.
\cite{LAZ95}.

We are therefore going to assume in the following that parameter $J$
describing these effects can be set to zero, or alternatively that its
effects can be mimicked redefining the U parameter as $U_{eff}= U - J $,
a practice that have been sometime used in the literature \cite{Dudarev98}. 

The Hubbard correction to the energy functional, Eqs. \ref{ub1} and
\ref{ub2}, greatly simplifies and reads:
\begin{eqnarray}
\label{our1}
E_{U}[\{n^{I\sigma}_{mm'}\}] &=& E_{Hub}[\{n^{I}_{mm'}\}] -
E_{dc}[\{n^{I}\}] \nonumber \\
&=&\frac{U}{2}\sum_{I}\sum_{m,\sigma}
\{n_{mm}^{I\sigma}-\sum_{m'}n_{mm'}^{I\sigma}n_{m'm}^{I\sigma}\}
\nonumber \\
&=&\frac{U}{2}\sum_{I,\sigma}
Tr[{\bf n}^{I\sigma}(1-{\bf n}^{I\sigma})].
\end{eqnarray}

Choosing for the localized orbitals the representation that diagonalizes 
the occupation matrices
\begin{eqnarray}
\label{diago}
{\bf n}^{I\sigma} {\bf v}^{I\sigma}_i = 
\lambda^{I\sigma}_i {\bf v}^{I\sigma}_i
\end{eqnarray} 
with $0 \le \lambda^{I\sigma}_i \le 1$, the energy correction becomes
\begin{eqnarray}
\label{our2}
E_{U}[\{n^{I\sigma}_{mm'}\}] 
&=&\frac{U}{2}\sum_{I,\sigma} \sum_{i} \lambda^{I\sigma}_i ( 1 -\lambda^{I\sigma}_i ).
\end{eqnarray}
from where it appears clearly that the energy correction introduces a
penalty, tuned by the value of the U parameter, for partial occupation
of the localized orbitals and thus favors disproportionation in fully
occupied ($\lambda \approx 1$) or completely empty ($\lambda \approx
0$) orbitals.
This is the basic physical effect built in the aDFT+U functional and
its meaning can be traced back to known deficiencies of LDA or GGA for
atomic systems.

\begin{figure}[!t]
\includegraphics[width=9.0truecm]{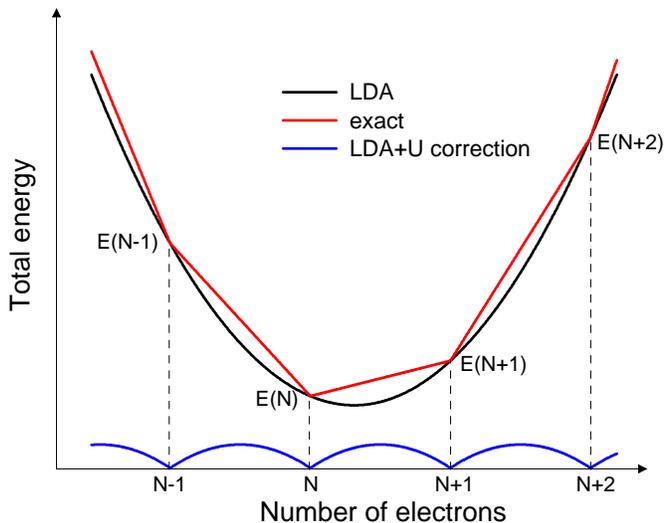}
\caption{\label{parab} Sketch of the total energy profile as a function of 
number of electrons in a generic atomic system in contact with a reservoir.
The bottom curve is simply the difference
between the other two (the LDA energy and the "exact" result
for an open system).}

\end{figure}

An atom in contact with a reservoir of electrons can exchange integer numbers
of particles with its environment. 
The intermediate situation with fractional number of electrons
in this open atomic system is described not by a pure state 
wave function, but rather by a statistical mixture so that, for instance, 
the total energy of a system with $N+\omega$ electrons (where N is an 
integer and $0 \le \omega \le 1$) is given by:
\begin{equation}
\label{efrac}
E_{n} = (1-\omega)E_{N} + \omega E_{N+1}
\end{equation}
where $E_{N}$ and $E_{N+1}$ are the energies of the system corresponding
to states with $N$ and $N+1$ particles respectively, while $\omega$
represents the statistical weight of the state with $N+1$ electrons.
The total energy of this open atomic system is thus represented by a
series of straight-line segments joining states corresponding to integer
occupations of the atomic orbitals as depicted in fig. \ref{parab}.
The slope of the energy vs electron-number curve is instead piece-wise
constant, with discontinuity for integer number of electrons, and corresponds
to the electron affinity (ionization potential) of the N (N+1) electron system.

Exact DFT correctly reproduce this behavior \cite{perd2,perd0},
which is instead not well described by the LDA or GGA
approach, which produces total energy with unphysical curvatures for
non integer occupation and spurious minima in correspondence of
fractional occupation of the orbital of the atomic system. This leads
to serious problems when one consider the dissociation limit of
hetero-polar molecules or an open-shell atom in front of a metallic
surface \cite{perd2,perd0}, and is at the heart of the 
LDA/GGA failure in the description of strongly correlated systems\cite{AZA91}. 
The unphysical curvature is associated basically to the incorrect
treatment by LDA or GGA of the self-interaction of the partially occupied
Kohn-Sham
orbital that gives a non-linear contribution to the total energy with
respect to orbital occupation 
(with mainly a quadratic term coming from
the Hartree energy not canceled properly in the exchange-correlation
term).

Nevertheless, it is well known \cite{jones1} that total energy differences
between different states can be reproduced quite accurately by the LDA (or GGA)
approach, if the occupation of the orbitals is {\it constrained} to assume
integer values.
As an alternative, we can recover the physical situation 
(an approximately piece-wise linear
total energy curve) by adding a correction to the LDA total energy which
vanishes for integer number of electrons and eliminates the curvature of
the LDA energy profile in every interval with fractional occupation
(bottom curve of fig. \ref{parab}). But this is exactly the kind of
correction that is provided by eq. \ref{our1} if 
the numerical value of the parameter $U$ 
is set equal to the curvature of the LDA (GGA) energy profile.

This clarifies the meaning of the interaction parameter $U$ as the
(unphysical) curvature of the LDA energy as a function of $N$ which is
associated with the spurious self-interaction of the fractional electron
injected into the system.
From this analysis it is clear that the numerical value of $U$ will
depend in general not only, as noted in the preceding section, on the
definition adopted for the occupation matrices but also on the particular
approximate exchange-correlation functional to be corrected, and should
{\it vanish} if the exact DFT functional were used. 

The situation is of course more complicated in solids where fractional
occupations of the atomic orbitals can occur due to hybridization of
the localized atomic-like orbitals with the crystal environment and
the {\it unphysical} part of the curvature has to be extracted from the
total LDA/GGA energy, which contains also hybridization effects. In the
next section this problem is discussed and a linear response approach
to evaluate Hubbard $U$ is proposed.

\subsection{Internally consistent calculation of U}

Following previous seminal works \cite{McMahan88,HSC89,AnGun91} we
compute $U$ by means of constrained-density-functional calculations
\cite{ConstrDFT84}.
What we need is the total energy as a function of the localized-level 
occupations of the "Hubbard" sites:

\begin{equation}
\label{eq:Eofn}
E [\{q_{I}\}] = \min_{n({\bf r}),\alpha_I} \left \{ 
E [n({\bf r})] + \sum_I \alpha_I ( n_{I} - q_{I} )\right \} ,
\end{equation}
where the constraints on the site occupations, $n_{I}$'s from Eq. \ref{eq:nI},
are applied employing the Lagrange
multipliers, $\alpha_I$'s. From this dependence we can compute numerically
the curvature of the total energy with respect to the variation, around the
unconstrained values $\{n^{(0)}_I\}$ , of the occupation
of one isolated site.
A supercell approach is adopted in which occupation of one 
representative site in a sufficiently large supercell 
is changed leaving unchanged all other site occupations.
This curvature contains the energy cost associated to 
the localization of an electron on the chosen site 
including all screening effects from the crystal environment,
but it is not yet the Hubbard $U$ we want to compute. 
In fact, had we computed the same quantity from the total energy of
the non-interacting Kohn-Sham problem associated to the same system,
\begin{equation}
E^{KS} [\{q_I\}] = \min_{n({\bf r}),\alpha_I} \left\{ E^{KS} [n({\bf r})] + 
\sum_I \alpha^{KS}_I ( n_I - q_I ) \right \},
\end{equation}
we would have obtained a non vanishing results as well because by
varying the site occupation a rehybridization of the localized orbitals 
with the other degrees of freedom is induced that gives rise to a non-linear 
change in the energy of the system.
This curvature coming from rehybridization, originating from the
non-interacting band structure but present also in the interacting case,
has clearly nothing to do with the Hubbard $U$ of the interacting system
and should be subtracted from the total curvature: 
\begin{equation}
\label{dedn1}
U = \frac{\partial^{2}E[\{q_I\}]}{\partial q_I^{2}} -
\frac{\partial^{2}E^{KS}[\{q_I\}]}{\partial q_I^{2}} .
\end{equation}

In Ref.\ \cite{AnGun91} Anisimov and Gunnarsson,
in order to avoid dealing with the above mentioned non-interacting curvature,
exploited the peculiarities of the LMTO method, used in their calculation, 
and decoupled the chosen localized orbitals from the remainder of the crystal
by suppressing in the LMTO hamiltonian the corresponding hopping terms.
This reduced the problem to the one of an isolated atom
embedded in an artificially disconnected charge background.
Thanks to Janak theorem \cite{Janak} the second order derivative
of the total energy in Eq.\ \ref{eq:Eofn} can then be recast as a first
order derivative of the localized-level eigenvalue.
In our approach the role played in Refs.\ \cite{AnGun91,SDA94} 
by the eigenvalue of the artificially isolated atom is taken by the 
Lagrange multiplier, used to enforce level occupation\cite{ConstrDFT84}:
\begin{eqnarray}
\label{deriv1}
&&\frac{\partial E[\{q_{J}\}]}{\partial q_{I}} = - \alpha_{I}, \quad 
\frac{\partial^{2}E[\{q_{J}\}]}{\partial q_{I}^{2}} = - \frac{\partial\alpha_{I}}{\partial q_{I}},\\
\nonumber 
&&\frac{\partial E^{KS}[\{q_{J}\}]}{\partial q_{I}} = - \alpha^{KS}_{I}, \quad
\frac{\partial^{2}E^{KS}[\{q_{J}\}]}{\partial q_{I}^{2}} = - \frac{\partial\alpha^{KS}_{I}}{\partial q_{I}}.
\end{eqnarray}
At variance with the original method of Refs.\ \cite{AnGun91,SDA94},
in our approach we need to compute and subtract the band-structure
contribution, $-{\partial\alpha^{KS}_{I}}/{\partial q_{I}}$, from
the total curvature, but, in return, Hubbard $U$ is computed in exactly the
same system to which it is going to be applied and the screening from
the environment is more realistically included. The present method was
inspired by the linear response scheme proposed by Pickett and coworkers
\cite{Pickett98} where however the role of the non-interacting curvature
was not appreciated.

In actual calculations constraining the localized orbital occupations is
not very practical and it is easier to pass, via a Legendre transform,
to a representation where the independent variables are the $\alpha_I$'s
\begin{eqnarray}
\label{eq:Eofa}
E [\{\alpha_I\}] & =& \min_{n({\bf r})} \left \{ E [n({\bf r})] 
 + \sum_I \alpha_I \; n_{I} \right\}, \\
\nonumber E^{KS}[\{\alpha^{KS}_I\}] & =& \min_{n({\bf r})} \left \{
E^{KS} [n({\bf r})] + \sum_I \alpha^{KS}_I n_{I} \right\}.
\end{eqnarray}
Variation of these functionals with respect to wavefunctions shows
that the effect of the $\alpha_I$'s is to add to the single particle
potential a term, $\Delta V = \sum_I \alpha_I P^{I} $ (or $ \Delta V
= \sum_I \alpha^{KS}_I P^{I} $ for the non-interacting case), where
localized potential shifts of strength $\alpha_I$ ($\alpha^{KS}_I$)
are applied to the localized levels associated to site $I$.

It is useful to introduce the (interacting and non-interacting) density
response functions of the system with respect to these localized perturbations:
\begin{eqnarray}
\label{chidef}
\chi_{IJ} 
& = & \frac{\partial ^2E}{\partial \alpha_I \partial \alpha_J} 
= \frac{\partial n_I}{\partial \alpha_J}, 
\\
\chi^0_{IJ} 
& = & \frac{\partial ^2E^{KS}}{\partial \alpha^{KS}_I \partial \alpha^{KS}_J} 
= \frac{\partial n_I}{\partial \alpha^{KS}_J}. 
\nonumber
\end{eqnarray}
Using this response-function language, the effective interaction
parameter $U$ associated to site $I$ can be recast as:
\begin{equation}
\label{ueff1}
U = + \frac{\partial \alpha^{KS}_I}{\partial q_I} -
\frac{\partial \alpha^I}{\partial q_I} = \left( \chi_{0}^{-1} -
\chi^{-1} \right)_{II}
\end{equation}
that is reminiscent of the well known random-phase approximation
 \cite{RPA} in linear response theory giving the interacting
density response in terms of the non-interacting one and the
Coulomb kernel. A similar result is obtained within DFT linear
response \cite{epsDFT} where the interaction kernel also contains an
exchange-correlation part.

The response functions, Eq.\ \ref{chidef}, needed in Eq.\ \ref{ueff1}
are computed taking numerical derivatives. We perform a well converged
LDA calculation for the unconstrained system ($\alpha_I=0$ for all sites
in the supercell) and---starting from its self-consistent potential---we
add small (positive and negative) potential shifts on each non equivalent
"Hubbard" site $J$ and compute the variation of the occupations, $n_I$'s,
for all sites in the supercell in two ways: $i)$ letting the Kohn-Sham
potential of the system readjust self-consistently to optimally screen
the localized perturbation, $\Delta V=\alpha_J P_J$, and $ii)$ without
allowing this screening. This latter result is nothing but the
variation computed from the first iteration in the self-consistent cycle
leading eventually to the former (screened) results. The site-occupation
derivatives calculated according to $i)$ and $ii)$ give the matrices 
$\chi_{IJ}$ and $\chi^0_{IJ}$ respectively.

\subsection{Further considerations}

Before moving to examine some specific examples in the next section,
let's end the present one discussing a few additional technical points.

As mentioned earlier, Hubbard $U$ is computed, ideally, from variation
of the site occupation of a single site in an infinite crystal and
in practice adopting a supercell approach where  periodically repeated
sites are perturbed coherently. In order to speed up the convergence
of the computed $U$ with supercell size it may result useful to enforce
explicitly charge neutrality for the perturbation, that is to be introduced
in the response functions, thus enhancing its local character and reduce 
the interaction with its periodic images.
In this procedure we introduce in the response functions, $\chi$ and
$\chi^0$, ---in addition to the degrees of freedom associated to the
localized sites---also a "delocalized background" representing all other
degrees of freedom in the system.
This translates in one more column and row in the response matrices, 
whose elements are determined imposing overall charge neutrality of the 
perturbed system for all localized perturbations, 
($\sum_I \chi_{IJ}  = 0, \quad \sum_I \chi^0_{IJ}=0, \quad \forall J$)
and absence of any charge density variation upon perturbing the system 
with a constant potential 
($\sum_J \chi_{IJ}  = 0, \quad \sum_J \chi^0_{IJ}=0, \quad \forall I$).
From a mathematical point of view both $\chi$ and $\chi_{0}$ acquire
a null eigenvalue, corresponding to a constant potential shift, and
the needed inversions in Eq.\ \ref{ueff1} must be taken with care. It
can be shown that their singularities cancel out when computing the
difference $\chi_{0}^{-1} - \chi^{-1}$ and the final result is well
defined.  We stress that in the limit of infinitely large supercell the
coupling with the background gives no contribution to the computed $U$,
but we found that this limit is approached more rapidly when 
this additional degrees of freedom is included.

In the same spirit we found that the spatial locality of the response
matrices can be rather different from the one of their inverse and a
supercell sufficient to decouple the periodically repeated response may
be too small to describe correctly the inverse in eq. \ref{ueff1}.  As a
practical procedure, therefore, after evaluating the response function
matrices in a given supercell, we extrapolate the result to much larger
supercells assuming that the most important matrix elements in $\chi_{0}$
and $\chi$ involve the atoms in the few nearest coordination-shells
accessible in the original supercell. The corresponding matrix elements
of the larger supercell are filled with the values extracted from the
smaller one while all other, more distant, interactions are neglected.
Again, when a sufficiently large supercell to extract the matrix elements
of the response functions is considered, 
the effect of this extrapolation vanishes, but,
as we will see in the following, this scheme capture a large fraction
of the system-size dependence of the calculated $U$ and it may allow to
reach more rapidly the converged result.

As a final remark we notice that the electronic structure of a system
described within the LDA+U approach may largely differ from the one
obtained within the LDA used to compute $U$. In a more refined
approach one might seek internal consistency between the band structure
used in the calculation of $U$ and the one obtained using it. We have not
addressed this issue here, but one can imagine performing the same type
of analysis leading to the $U$ determination for a functional already
containing an LDA+U correction. The computed $U$ would in that case be
a correction to be added to the original $U$ and internal consistency
would be reached when the correction vanishes.

\section{Examples}

\subsection{Metals: Iron and Cerium}

In their seminal paper Anisimov and Gunnarsson \cite{AnGun91} computed
the effective on site Coulomb interaction between the localized electrons
in metallic Fe and Ce. For Ce the calculated Coulomb interaction was
about 6 eV in good agreement with empirical and experimental estimates
ranging from 5 to 7 eV \cite{ConstrDFT84,Herbst78,ExpCeU}, while the
result for Fe (also about 6 eV) was surprisingly high since $U$ was
expected to be in the range of 1-2 eV for elemental transition metals,
with the exception of Ni \cite{ExpFeU1,ExpFeU2}. Let us apply the present
approach to these two system, starting with Iron.

In its ground state elemental Iron has a ferromagnetic (FM) spin
arrangement and a body-centered cubic (BCC) structure. Gradient corrected
exchange-correlation functional are needed in order to stabilize the
experimental structure as compared with non-magnetic face-centered cubic
(FCC) structure preferred by LDA. The Perdew-Burke-Ernzherof (PBE)
\cite{perd1} GGA functional was employed here. Iron ions were represented
by ultrasoft pseudopotential and kinetic energy cutoffs of 35 Ry and 420
Ry were adopted for wavefunction and charge density Fourier expansion.
Brillouin Zone integrations where performed using 8$\times$8$\times$8
Monkhorst and Pack special point grids \cite{AB} using Methfessel and
Paxton smearing technique \cite{met1} with a smearing width of
0.005 Ry in order to smooth the Fermi distribution.

The calculation of the effective Hubbard $U$ followed the procedure
outlined in preceding section: a supercell
was selected containing a number of inequivalent Iron atoms; then,
after a well converged self-consistent calculation, we applied to one
of these atoms small, positive and negative, potential shifts, $\Delta
V= \alpha P_{d}$ (with $\alpha = \pm $0.2-0.5 eV), where  $P_d$ is the
projector on the localized $d$ electron of the selected atom. From the
variation of the $d$-level occupations of all Iron atoms in the cell
one column of $\chi$ and $\chi^0$ response functions was extracted and
all other matrix elements were reconstructed by symmetry, including
the background as explained previously. Hubbard $U$
was then calculated from Eq.~\ref{ueff1}.

In order to describe response for an isolated perturbation four
supercells were considered: {\it i)} a simple cubic (SC) cell containing
two inequivalent iron atoms, the perturbed atom and one of its nearest
neighbors; {\it ii)} a 2$\times$2$\times$2 BCC supercell containing
8 inequivalent Iron atoms, 4 in the nearest-neighbor shell of the
perturbed atom and 3 belonging to the second shell of neighbors; {\it
iii)} a 2$\times$2$\times$2 SC cell containing 16 atoms, including also
some third nearest-neighbor atom and {\it iv)} a 4$\times$4$\times$4
BBC supercell containing 64 inequivalent Iron atoms; we used this largest
cell just to extrapolate the results from the smaller ones. 

\begin{figure}[!t]
\includegraphics[width=9.5truecm]{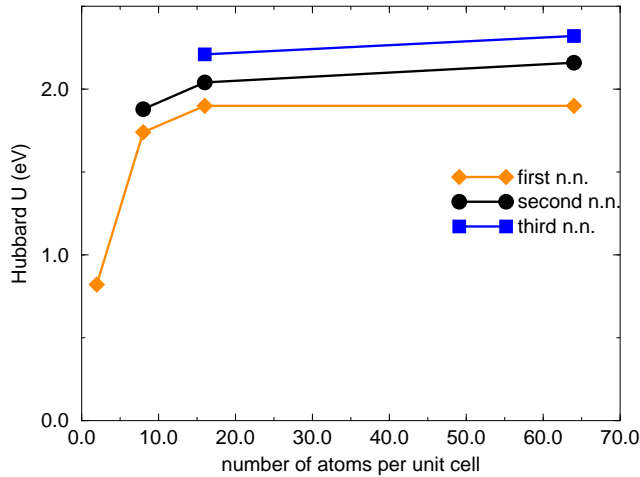}
\caption{Calculated Hubbard $U$ in metallic Iron for different
supercells. Lines connect results from the cell-extrapolation
procedure described in the text and different symbols correspond to inclusion
of screening contributions up to the indicated shell of neighbors of
the perturbed atom.}

\label{unei} 
\end{figure}

The convergence properties of the effective $U$ of bulk iron with the size
of the used supercell are shown in fig.~\ref{unei}.  

\begin{figure}[!t]
\includegraphics[width=9.5truecm]{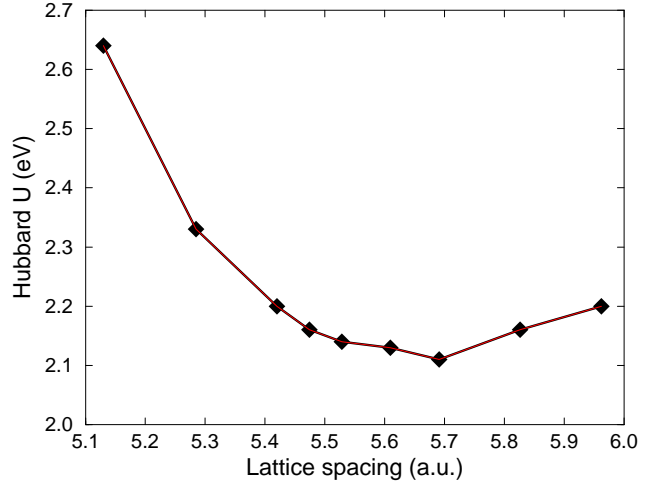}
\caption{Lattice spacing dependence of the calculated Hubbard $U$
parameter for Iron.}
\label{uls}
\end{figure}

The Hubbard $U$ obtained from the SC 2-atom cell, once inserted in the
64-atom supercell, captures most of the effective interaction; second
nearest neighbors shell brings some significant corrections to the final
extrapolated result, while third nearest neighbor shell has a smaller
effect. We believe that contributions from further neighbor rapidly
vanish and that an accurate value of $U$ can be extracted from the SC
supercell containing 16 atoms. The extrapolation from this cell to
larger cells brings only minor variations which are within the finite
numerical accuracy that we estimate within a fraction of an eV. From
this analysis our estimate for the Hubbard $U$ in elemental Iron at the
experimental lattice parameter is therefore 2.2 $\pm$ 0.2 eV.

This results is in very good agreement with the experimental estimates
\cite{ExpFeU1,ExpFeU2}, but disagrees with Anisimov and Gunnarsson
result \cite{AnGun91}. We can only recall here that many technical
details differ in the two approaches. In particular {\it i)} in the
original approach the perturbed atom is disconnected from the rest of the
crystal by removing all hopping terms, thus leaving a rather unphysical
environment to perform the screening, while in our approach the actual
system is allowed to screen the perturbation, {\it ii)} the Atomic Sphere
Approximation (ASA) was employed in the original LMTO calculation while
no shape approximation is made in our case.

In order to further test our approach on this element we investigate the
dependence of the Hubbard parameter on crystal structure. The dependence
of the calculated interaction parameter on the lattice spacing of
the unit cell is shown in fig.~\ref{uls} where a marked increase of
the Hubbard $U$ can be observed when the lattice parameter is squeezed
below its experimental value. Despite this may appear counterintuitive,
as correlation effects are expected to become less important when atoms
gets closer, one should actually compare the increasing value of $U$
with the much steeper increase of bandwidth when reducing the interatomic
distance. Upon increase of the lattice parameter the Hubbard parameter
should approach the atomic limit that can be estimated from all-electron
atomic calculations where the local neutrality of the metallic system
is maintained: $U = E(d^8s^0) + E(d^6s^2) - 2 \times E (d^7s^1) = 2.1 $
eV, in reasonable agreement with the results of fig.~\ref{uls}.

Using the calculated volume dependent Hubbard $U$ parameter we have studied 
the effect of the LDA+U approximation on the structural properties of Iron.
\begin{table}[!t]
\caption{ Comparison between the calculated lattice constant ($a_{0}$), 
bulk modulus ($B_{0}$) and magnetic moment ($\mu_{0}$) within several 
approximate DFT schemes and experimental results quoted from
\protect\cite{exp1}. }
\label{ulats}
\vspace{0.6truecm}
\begin{tabular}{lccc}
& $a_{0}$ (a.u.)&  $B_{0}$ (Mbar)& $\mu_{0}$ ($\mu_{B}$)\\
 \hline
Expt.        & 5.42 & 1.68 & 2.22 \\
             &      &      &      \\
LSDA         & 5.22 & 2.33 & 2.10 \\
$\sigma$-GGA & 5.42 & 1.45 & 2.46 \\
LDA+U        & 5.53 & 2.12 & 2.60 \\
LDA+U (AMF)  & 5.34 & 1.53 & 2.00 \\
\end{tabular}
\end{table}
Results  are reported in table \ref{ulats}  where they are compared with results
obtained within LSDA and  $\sigma$-GGA(PBE) approximation and with
experimental data. From these data it appears that, although simple
$\sigma$-GGA(PBE) approximation appears to be superior in this case,
LDA+U provides a reasonable description of the data, of the same quality
as LSDA. In weakly correlated metals it has been suggested \cite{amf}
that a formulation of LDA+U in terms of occupancy fluctuations around
the uniform occupancy of the localized level could be more appropriate
than the standard one.
This "around mean field" (AMF) LDA+U approach has been revisited recently
\cite{amf1,amf2} and an "optimally mixed" scheme has also been proposed 
\cite{amf2}. We don't want to enter in this discussion here, but  
we mention that by following the AMF recipe the
description of structural and magnetic properties of metallic
Iron improves as it is evident from table \ref{ulats}.

\begin{figure}[t]
\includegraphics[width=9.0truecm]{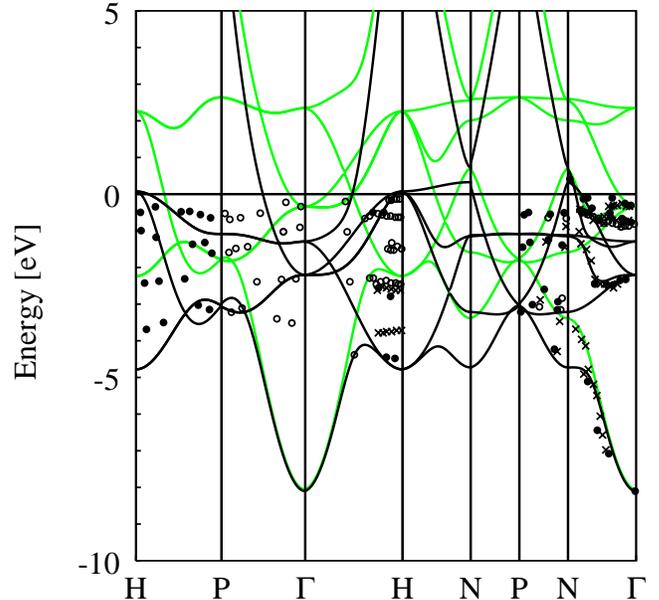}
\caption{Band structure of bulk iron obtained
within the AMF LDA+U approach. Green lines are for minority
spin states, black ones for majority spin levels.
Photoemission results from \protect\cite{turner1} are
also reported for comparison.}
\label{febamf}
\end{figure}

Using the calculated value of $U$ we have obtained the electronic
structure of Iron at the experimental lattice spacing. The theoretical
band structure obtained using the AMF version of LDA+U is reported
in fig.\ \ref{febamf} together with some experimental results \cite{turner1}.
The overall agreement is rather good for this scheme. However, when using
the standard LDA+U scheme a somehow worse agreement with experimental
data was obtained, mainly due to a rigid downward shift of the majority
spin bands of about 1 eV. This is an indication that LDA+U approximation
may still require some fine tuning in order to describe accurately both
strongly and weakly correlated systems \cite{amf2}.

Let us proceed to examine the Cerium case.  
Elemental cerium presents a very interesting phase diagram with a peculiar
isostructural $\alpha-\gamma$ phase transition between a low volume
($\alpha$) and a high volume ($\gamma$) phase, both FCC.
This phase transition has attracted much experimental and 
theoretical interest and in the last 20 years \cite{cerium_varie},
many interpretations have been put forward to explain its occurrence.
It is clear now that standard LDA or GGA approximations do not describe the
transition and it appears that a treatment of the correlation 
at the DMFT level might be required \cite{cerium_DMFT}, however 
a full understanding of the nature of the transition 
is still under debate \cite{cerium_last_prl}.
Here, we do not want to address this delicate topics but we simply want
to follow  Anisimov and Gunnarsson \cite{AnGun91} by computing the Hubbard
$U$ parameter for elemental cerium in the high volume $\gamma$ phase.

The interaction of valence-electrons with Ce nuclei and its core electrons
was described by a non-local ultrasoft pseudopotential \cite{USPP}
generated in the $5s^25p^65d^14f^1$ electronic configuration. Kinetic
cutoffs of 30 Ry and 240 Ry were adopted for wavefunction and charge
density Fourier expansion. The LSDA approximation was adopted for the
exchange and correlation functional. Brillouin Zone integrations where
performed using 8$\times$8$\times$8 Monkhorst and Pack special point grids
\cite{AB} using Methfessel and Paxton smearing technique \cite{met1} with
a smearing width of 0.05 Ry.

\begin{figure}[t]
\includegraphics[width=8.0truecm]{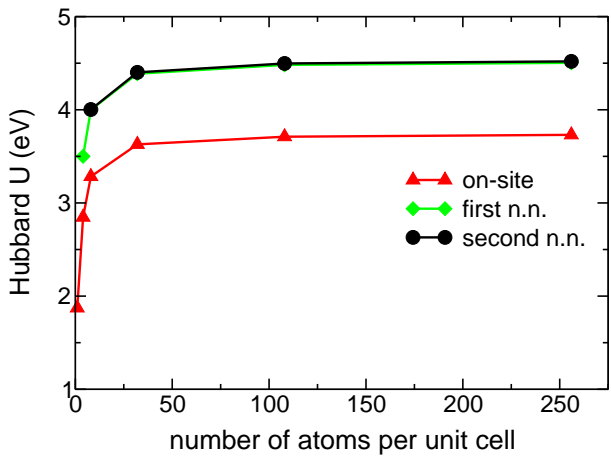}
\caption{Calculated Hubbard $U$ in metallic Cerium for different
supercells. Lines connect results from the cell-extrapolation procedure
and different symbols correspond to inclusion of screening contributions
up to the indicated shell of neighbors of the perturbed atom.}
\label{fig:CeU}
\end{figure}
To obtain the response to an isolated perturbation we have perturbed
a Cerium atom in three different cells: {\it i)} the fundamental
face-centered cubic (FCC) cell containing just one inequivalent atom, {\it
ii)} a simple-cubic (SC) cell containing 4 atoms (giving access to the
first nearest-neighbor response) and {\it iii)} a 2$\times$2$\times$2 FCC
cell (8 inequivalent atoms) including also the response of second-nearest
neighbor atoms.  The result of these calculations and their extrapolation
to very large SC cells is reported in Fig.\ \ref{fig:CeU} where it can
be seen that the converged value for $U$ approaches 4.5 eV.

The screening in metallic cerium is extremely localized, as can be seen
from the fact that inclusion of the first-nearest neighbor response
is all is needed to reach converged results. This is at variance with
what we found in metallic Iron where third nearest-neighbor response
was still significant (see Fig.\ \ref{unei}). The calculated value is
not far from the value (5-7 eV) expected from empirical and experimental 
estimates \cite{ConstrDFT84,Herbst78,ExpCeU}, especially if we consider that
the parameter $U$ we compute plays the role of $U-J$ in the simplified
rotational invariant LDA+U scheme adopted \cite{Dudarev98}. 

As a check, we performed all-electron atomic calculations for $Ce^+$ ions
where localized $4f$ electrons were promoted to more delocalized $6s$ or
$5d$ states and obtained $U = E(f^3s^0) + E(f^1s^2) - 2 \times E (f^2s^1)
= 4.4 $ eV, or $U = E(f^2s^0d^1) + E(f^0s^2d^1) - 2 \times E(f^1s^1d^1)
= 6.4$ eV, depending on the selected atomic configurations. This confirms
the correct order of magnitude of our calculated value in the metal.

The present formulation is therefore able to provide reasonable values for
the on-site Coulomb parameter both in Iron and Cerium, at variance with
the original scheme of ref.\ \cite{AnGun91} where only the latter
was satisfactorily described. We believe that a proper description of
the interatomic screening, rather unphysical in the original scheme where
atoms were artificially disconnected from the environment, is important
to obtain a correct value for Hubbard $U$ parameter, especially in Iron
where this response is more long-ranged.

\subsection{Transition metal monoxides: FeO and NiO}

The use of the LDA+U method for studying FeO is mainly motivated by
the attempt to reproduce the observed insulating behavior.  In fact,
as for other transition metal oxides (TMO), standard DFT methods, as LDA
or GGA, produce an unphysical metallic character due to the fact that
crystal field and electronic structure effects are not sufficient in this
case to open a gap in the three-fold minority-spin $t_{2g}$ levels that
host one electron per Fe$^{2+}$ atom. As already addressed in quite abundant
literature on TMO (and FeO in particular), a better description of the
electronic correlations is necessary to obtain the observed insulating
behavior and the structural properties of this compound at low pressure
\cite{isa1,zon1,zon2,maz1}. The application of our approach to this
material will thus allow us to check its validity by comparison of our
results with the ones from experiments and other theoretical works.

\begin{figure}[!t]
\includegraphics[width=8.0truecm]{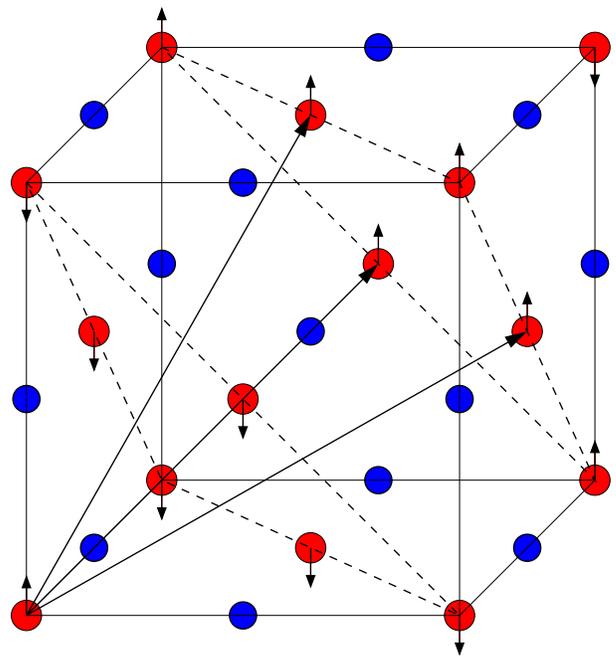}
\caption{\label{cell}The unit cell of FeO: 
blue spheres represent Oxygen ions, red ones are Fe ions, with
arrows showing the orientation of their magnetic moments. 
Ferromagnetic (111) planes of iron ions alternate with opposite spins 
producing type II antiferromagnetic order and rhombohedral symmetry.}
\end{figure}

The unit cell of this compound is of rock-salt type, with a rhombohedral
symmetry introduced by a type II antiferromagnetic (AF) order (see
fig. \ref{cell}) which sets in along the [111] direction below a Ne\'el
temperature of 198 K, at ambient pressure.

The calculations on this materials were all performed in the 
antiferromagnetic phase starting from the cubic (undistorted) unit
cell of fig. \ref{cell} with the experimental lattice spacing.
We used a 40 Ry energy cut off for the 
electronic wavefunctions (400 Ry for the charge density due to
the use of ultrasoft pseudopotentials \cite{USPP} both for Fe and O) and
a small smearing width of 0.005 Ry which required 
a 4$\times$4$\times$4 k-points mesh.

\begin{figure}[!t]
\includegraphics[width=9.0truecm]{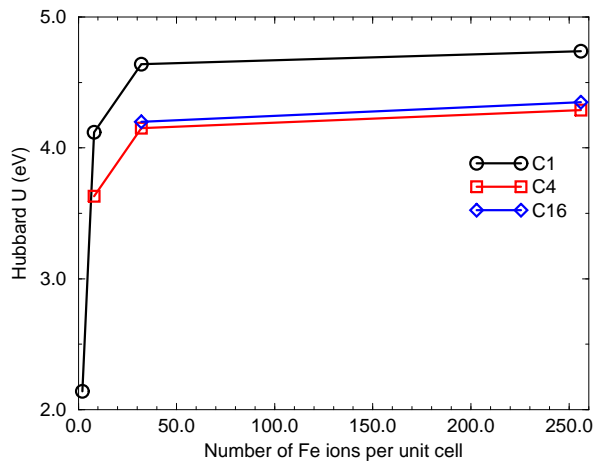}
\caption{\label{ufeoconv} Convergence of Hubbard $U$ parameter of FeO with
the number of iron included in the supercell used in the extrapolation.
Lines connect results including the screening contributions extracted
from the indicated cell.}
\end{figure}

To compute the Hubbard effective interactions, we performed GGA
calculations with potential shifts on one Hubbard site in larger and
larger unit cells, that we named C1, C4, and C16, containing 2, 8,
and 32 iron ions respectively, and extrapolated their results up to a
supercell containing 256 magnetic ions (called C128). The result for the
undistorted cubic cell at the experimental lattice spacing is reported
in fig. \ref{ufeoconv}. We can observe that the effective interaction
obtained from C4 is already very well converged, when extrapolated to
the largest cell, with respect to inclusion of screening from additional
shells of neighborers,

\begin{figure}[!t]
\includegraphics[width=8.5truecm]{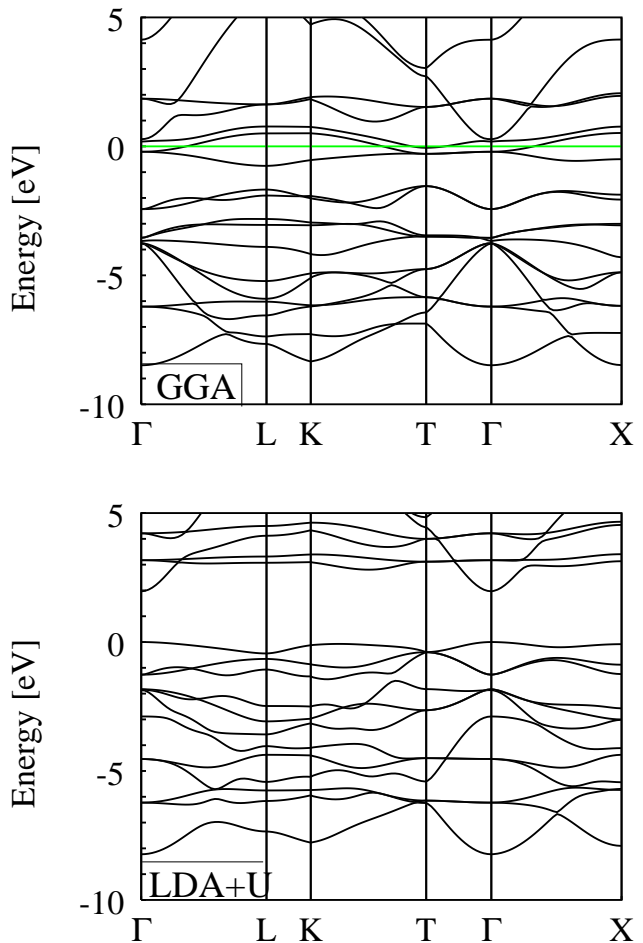}
\caption{\label{feob} The band structure of FeO in the undistorted (cubic) AF 
configuration at the experimental lattice spacing obtained within GGA (top
panel) and LDA+U using the computed Hubbard $U$ of 4.3 eV (bottom panel). 
The zero of the energy is set at the top of the valence band.}
\end{figure}

The final result for the Hubbard $U$ is 4.3 eV which
is smaller than most of the values obtained (or simply
assumed) in other works \cite{zon1,zon2,maz1}. 
If we use this value in a LDA+U calculation we can obtain 
the observed insulating behavior as shown in the band
structure plot of fig. \ref{feob}
where a comparison is made with GGA (metallic) results.

\begin{figure}[!h]
\includegraphics[width=8.5truecm]{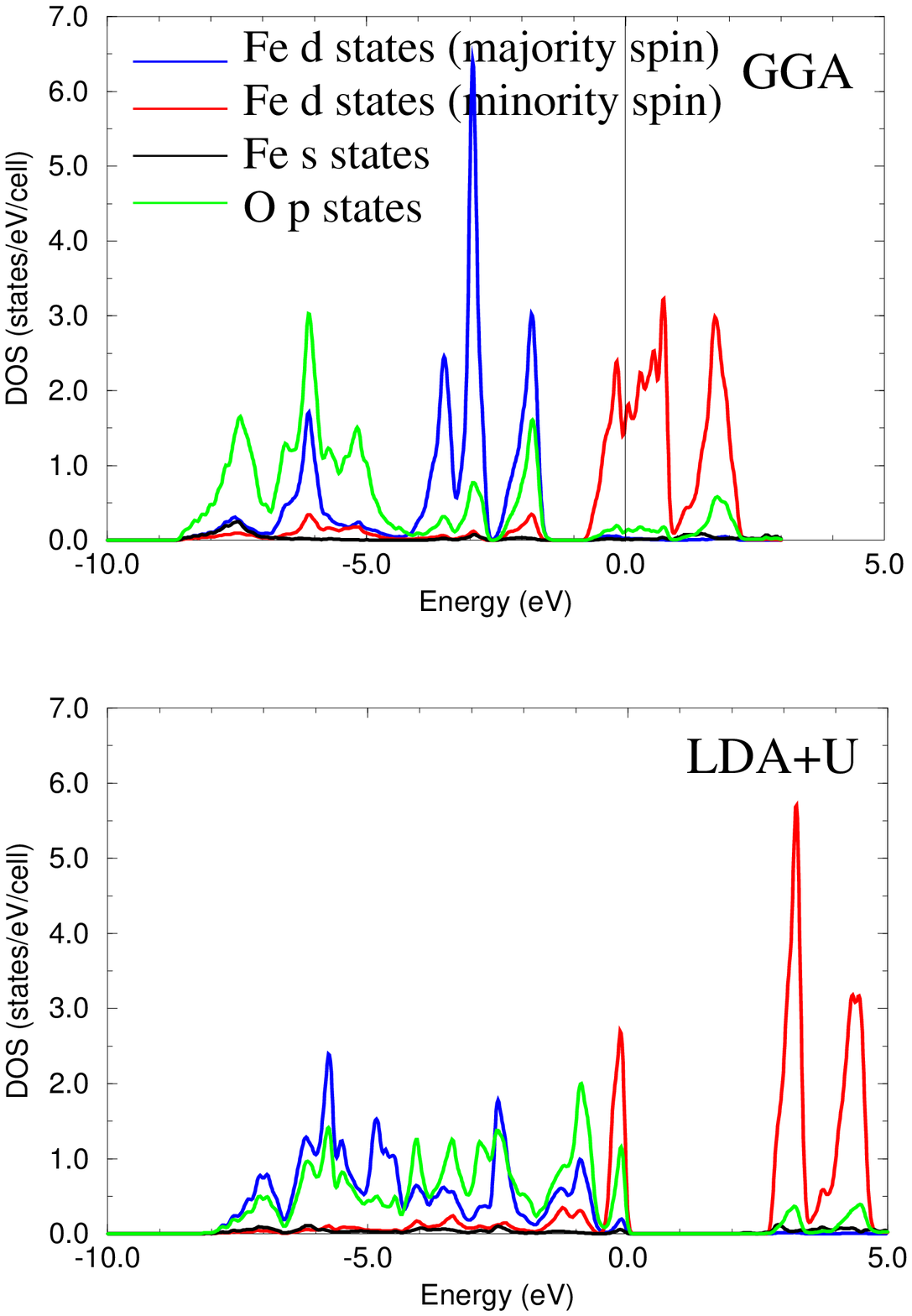}
\caption{\label{feodosu}
Projected density of states of FeO in the undistorted (cubic) AF
configuration at the experimental lattice spacing obtained within GGA (top 
panel) and LDA+U using the computed Hubbard $U$ of 4.3 eV (bottom panel).}
\end{figure}

A gap opens around the Fermi level whose minimal width is about 2 eV. The
band gap is direct and located at the $\Gamma$ point.  The corresponding
transition, of $3d$(Fe)-$2p$(O)$\rightarrow$$4s$(Fe) character, should
be quite weak due to the vanishing weight of Iron $s$ states at the
bottom of the valence band (fig. \ref{feodosu}, bottom picture). We can
expect that a stronger absorption line will appear instead around 2.6
eV due to the transition, of $3d$(Fe)-$2p$(O) $\rightarrow$ $3d$(Fe)
character, among two pronounced peaks of the density of states around
the Fermi level.  This picture is in very good agreement with experiments
(and other theoretical results \cite{maz1,wei1}) where a first weak
absorption is reported between 0.5 and 2 eV and a stronger line appears
around 2.4 eV \cite{jmmm}. The large mixing between majority-spin Iron
$3d$ states and the Oxygen $2p$ manifold over a wide region of energy
and the finite contribution of the Oxygen states at the top of the
valence band---a feature not present within $\sigma$-GGA (see top panel
in fig. \ref{feodosu})--- are also in good agreement with experiments,
which indicate for FeO a moderate charge transfer character of the
insulating state.

Despite our $U$ is smaller than the ones used in literature, we find a
good agreement of our results about the electronic structure of the system
with experiments and other theoretical works. These findings confirm
the validity of our internally consistent method to compute $U$. We now
want to extend its application to the study of structural properties.
This is indeed a very important test because a good ab-initio method
should be able to describe the true ground state of a system and provide
a complete description of both electronic and structural properties.
Furthermore the plane-wave implementation we use allows a straightforward
calculation of Hellmann-Feynman forces and stresses, thus giving easily
access to equilibrium crystal structure.

As observed in experiments \cite{yagi1}, the cubic rock salt structure
of FeO shown in fig. \ref{cell} becomes unstable under a pressure of
16 GPa (at room temperature) toward a rhombohedral distortion.  In the
distorted phase the unit cell is elongated along the [111] direction
with a consequent shrinking of the interionic distances on the (111)
planes. This transition is driven by the onset of the AFII magnetic
order \cite{yagi1} (the Ne\'el temperature reaches room value at about
16 GPa) which imposes a rhombohedral symmetry even in the cubic phase.
Upon increasing pressure above the threshold value the distortion of the
unit cell is observed to increase producing more elongated structures
\cite{yagi1}.

\begin{figure}[!t]
\includegraphics[width=8.0truecm]{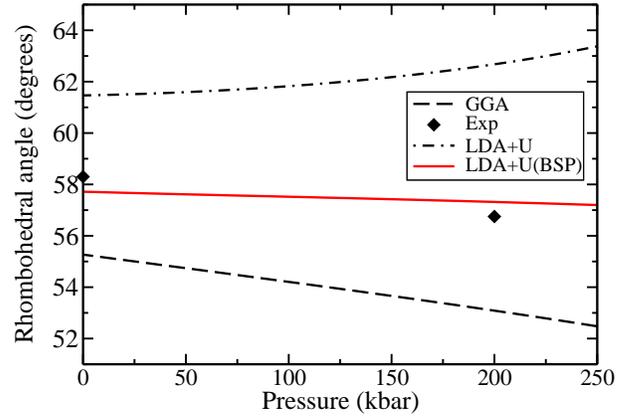}
\caption{\label{angp} The pressure dependence of the rhombohedral
angle in FeO for the various approximations described in the text is
compared with experimental results. These latter results were extracted
extrapolating the data for the non stoichiometric compound Fe$_{1-x}$O
up to the stoichiometric composition \cite{yagi1,will1}.}

\end{figure}

We have computed the Hubbard $U$ on a grid of possible values for the
rhombohedral distortion and cell parameter and then from the corresponding
total energy calculations we determined the rhombohedral distortion and
the enthalpy of the system as a function of the pressure up to 250 Kbar.

As evident from fig.\ \ref{angp}, while GGA overestimates the rhombohedral
distortion and his pressure dependence, LDA+U method---in the standard
electronic configuration examined so far--overcorrects the GGA
results and introduces even larger errors with respect to experimental
results. In fact not only we obtain a distortion with the wrong sign
(of compressive character along the [111] direction), but also the wrong
pressure dependence.  The reason for this failure can be traced back to
the different occupation of the orbitals around the gap/Fermi level in
the two cases. Even in the undistorted cell, the rhombohedral symmetry,
induced by the antiferromagnetic order, lifts the degeneracy of the
minority spin $t_{2g}$ states of iron and split them in one state of
$A_{1g}$ character---which is essentially the m=0 ($z^{2}$) state along
the [111] quantization axis---and two states of $e_g$ symmetry localized
on the iron (111) planes. Within GGA, the Iron minority-spin $3d$ 
electrons partially occupy the two equivalent $e_g$ orbitals giving rise to two
half filled bands and a (wrong) metallic state which is delocalized on the 
(111) plane. The system gains energy by filling
the lowest half of the $e_g$ states and tends to elongate in the [111]
direction, shrinking in the plane, because this increases the overlap of
the $e_g$ states and their bandwidth.  Within LDA+U, fractional occupation
of orbitals is energetically disfavored and the system would like to
have completely filled or empty $3d$ states. In the standard unit-cell
considered so far in the literature---and used by us in the calculation
above---this can be accomplished only by filling the non-degenerate
$A_{1g}$ level, corresponding to wavefunctions elongated along [111], and
pushing upward in energy the in-plane $e_g$ states, leaving them empty. As
a consequence, the system tends to pull apart the ions on the same (111)
plane, so that the bandwidth of the state in the plane is reduced, and
increases instead the inter-plane overlap of the $A_{1g}$ states. This
simple picture gives an explanation of the fact that GGA overestimates
the elongation of the unit cell in the [111] direction, as well as the
(wrong) compressive behavior of the standard LDA+U solution.
We are thus left with the paradoxical situation that a correct pressure
dependence of the structural properties can be obtained from the wrong
band structure and viceversa.

\begin{figure}[!t]
\includegraphics[width=8.0truecm]{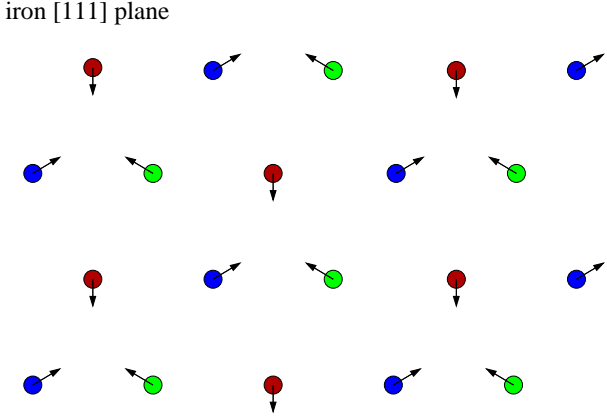}
\caption{\label{dist}
Lattice distortion in the (111) iron planes used to induce symmetry 
breaking in the electronic configuration of FeO.}
\end{figure}

We have found that it is possible to solve this paradox by allowing
the possibility that the system partially occupies, as within GGA,
the $e_g$ levels, thus maintaining the driving force for the right
rhombohedral deformation, and still opens a gap, as in standard LDA+U,
by some orbital ordering that breaks the equivalence of the iron ions
in the (111) plane. This possibility has been sometimes proposed in
literature \cite{maz1,cohen} but has never been clearly addressed.

From a simple tight-binding picture one finds that the optimal broken
symmetry phase would be the one where occupied $e_g$ orbitals have
the highest possible hopping term with unoccupied $e_g$ orbitals in
nearest-neighbor atoms in the plane, in order to maximize the kinetic
energy gain coming from delocalization, and the lowest possible hopping
term with neighboring occupied $e_g$ orbitals, in order to minimize
bandwidth that tends to destroy the insulating state. In bipartite lattice
this is simply achieved by making occupied orbitals in nearest-neighbor
sites orthogonal but, in the triangular lattice, formed by iron atoms
in (111) planes, this is not exactly possible, the system is topologically 
frustrated and some compromise is necessary.

It is generally believed \cite{HEISENBERG} that Heisenberg model in the
triangular lattice, to which our system resemble in some sense, displays a
three-sublattice 120$^\circ$ N\'eel long-range order.  We thus imposed a
symmetry breaking to the  system where three nearest-neighbor atoms in the
(111) plane were made inequivalent by slightly displacing them from the
ideal positions in the way shown in fig.\ \ref{dist}. This induced the
desired symmetry breaking of the electronic structure and opened a gap
that was robust and persisted when the atoms were brought back into the
ideal positions. We found, quite satisfactorily, that the new broken
symmetry phase (BSP) corresponds to a lower energy minimum than the
"standard" LDA+U solution and that therefore it is, to say the least, a
more consistent description of the ground state of FeO. The one depicted
in fig. \ref{dist} is, of course, only one of three equivalent distortions
we could have imposed to the electronic structure of the system and three
symmetry related BSPs could be defined. In the actual system an effective
equivalence of the ions in the (111) planes is probably restored by a
(dynamical) switching among equivalent states but considering the atoms
as strictly equivalent, as in the standard solution, leads to incorrect 
results.

\begin{figure}[!t]
\includegraphics[width=10.0truecm]{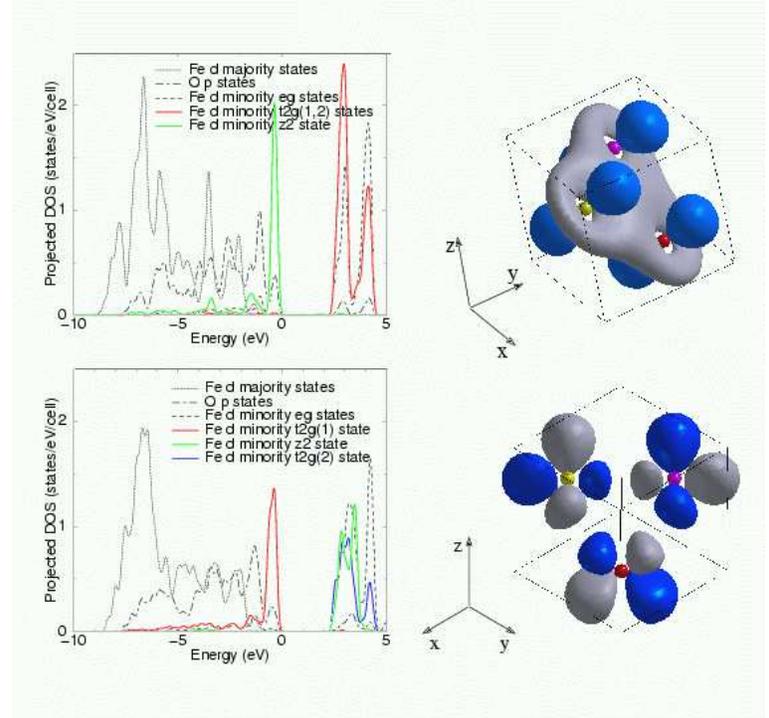}
\caption{\label{newdos} The projected density of states of FeO as obtained
in the "standard" LDA+U ground state (top panel) and in the proposed
broken symmetry phase (bottom panel). On the right of each DOS is a 
picture of the corresponding occupied Fe-$3d$ minority states.  }
\end{figure}

The comparison of the projected density of state in the "standard" LDA+U
solution and in the novel BSP phase is shown in fig.\ \ref{newdos} where
also a pictorial representation of the occupied minority-spin orbitals
in the two cases is shown. As we can observed, no remarkable qualitative
difference in the DOS appears apart from the different ordering of the
$d$ states around the gap. In fact the minority-spin $d$ electron is
now accommodated on a state lying on the (111) plane (shown on the right
panel) while the one with $A_{1g}$ ($z^{2}$) character has been pushed
above the energy gap. The gap width and the charge transfer character
of the system do not change significantly and are still in very good
agreement with the experiments.

We repeated the structural calculations (according to the same procedure
described above) in the BSP, and obtained the LDA+U (BSP) curve reported
in fig. \ref{angp}. The agreement with experiments is much improved with
respect to both GGA and LDA+U "standard" ground states.  The mechanism
leading to the pressure behavior in BSP case is basically the same already
producing the correct evolution of distortion in the GGA calculations.
When the unit cell elongates along the cubic diagonal the iron ions in the
(111) plane get closer and the hopping between nearest-neighbor orbitals
increased with a consequent lowering of the electronic kinetic energy.

We therefore conclude that LDA+U, not only improves the description of the
structural and electronic properties with respect to GGA, but that a close
examination of both electronic and structural properties is in this case
necessary in order to describe the correct ground state of the system.


Another classical example of TMO we want to study in order to test
the present implementation of LDA+U is Nickel Oxide. It is a very
well studied material and there is a good number of theoretical
\cite{Bengone00} and experimental works, including some photoemission
experiments \cite{zx1,kule1}, our results can be compared with. At
variance with FeO, no compositional instability is observed for NiO so
that the stoichiometric compound is easy to study and is much better
characterized than iron oxide. It has cubic structure with the same
AF spin arrangements of rhombohedral symmetry as FeO, but does not show
tendencies toward geometrical distortions of any kind and is therefore
easier to study.

In this case we did not perform any structural relaxation and calculated
the value of $U$ at the experimental lattice spacing for the cubic unit
cell imposing the rhombohedral AF magnetic order which is the ground state
spin arrangement for this compound.  The GGA approximation (in the PBE
prescription) was used in the calculation. US pseudopotentials for Nickel
and Oxygen (the same as in FeO) were used with the same energy cutoffs
(of 40 and 400 Ry respectively) for both the electronic wavefunctions
and the charge density as for FeO and also the same 4$\times$4$\times$4
k-point grid for reciprocal space integrations.

In the calculation of the Hubbard $U$ of NiO we did not studied the
convergence properties of $U$ with system size as we did in FeO but,
assuming a similar convergence also in this case, we performed a
constrained calculation only in the C4 cell and then extrapolated the
obtained result to the C128 supercell. The calculated value of the $U$
parameter is 4.6 eV. This value is smaller than literature values for
the same parameter that are rather in the range of 7-8 eV \cite{AZA91},
however it has been recently pointed out \cite{Dudarev98,Bengone00} that
in obtaining these values self-screening of $d$ electrons is neglected
and that better agreement with experimental results is obtained using an 
effective Hubbard $U$ of the order of 5-6 eV.

The magnetic moment of the Ni ions is correctly described within the
present GGA+U approach which gives a value of 1.7 $\mu_{B}$ well within
the experimental range of values ranging from 1.64 and 1.9 $\mu_{B}$
\cite{Alperin62,Cheetham83}, better than the value of 1.55 $\mu_{B}$
obtained within GGA. 

\begin{figure}[!t]
\includegraphics[width=8.5truecm]{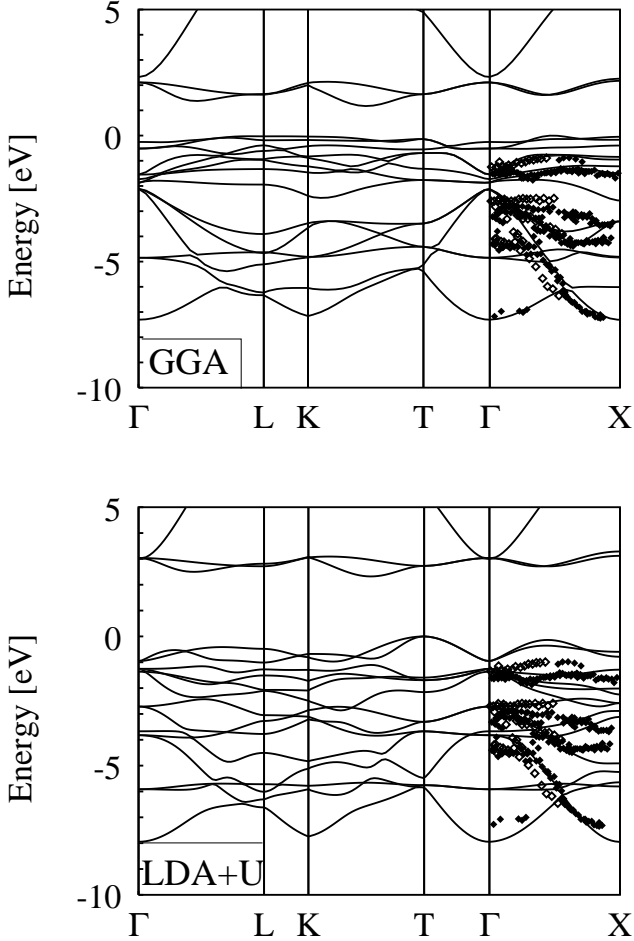}
\caption{\label{nioband}
The band structure of NiO in the undistorted (cubic) AF configuration at
the experimental lattice spacing obtained within GGA (top panel) and with
the computed Hubbard $U$ of 4.6 eV (bottom panel). The zero of the energy is
set at the top of the valence band. Experimental data from
ref.\  \protect\cite{zx1} (empty symbols) and \protect\cite{kule1}
(solid symbols) are also reported.}
\end{figure}

\begin{figure}[!t]
\includegraphics[width=8.7truecm]{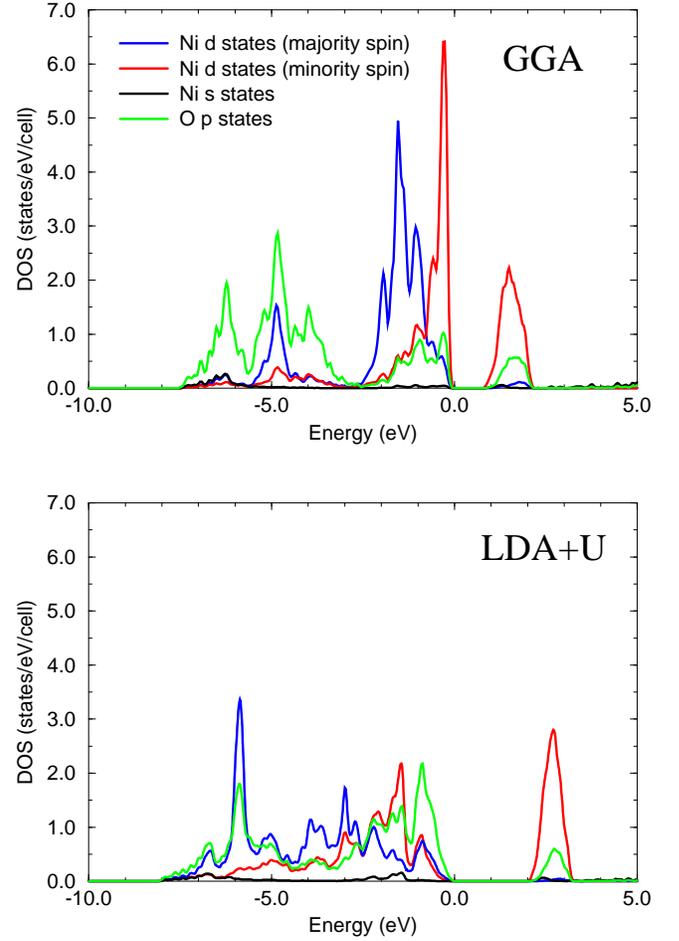}
\caption{\label{niodos}
Projected density of states of NiO in the undistorted AF configuration
at the experimental lattice spacing obtained with $U$ = 4.6 eV.}
\end{figure}

In fig.\ \ref{nioband} and fig.\ \ref{niodos} the band structure and
atomic-state projected density of states of NiO obtained with this value
of $U$ is shown, along with the results of standard GGA, and compared
with the photoemission data in the $\Gamma$X direction extracted from
ref. \cite{zx1,kule1}.

Despite the agreement with the experimental band-dispersion is not
excellent---the valence band width is somehow overestimated by both
GGA and GGA+U calculations---, GGA+U band structure reproduces well some
features of the photoemmission spectrum for this compound and gives a much
larger band gap than the one obtained within GGA approximation. A very
important feature to be noticed in the density of states reported in fig.\
\ref{niodos} is the fact that GGA+U modifies qualitatively the nature of
the states at the top of the valence band, and hence the nature of the
band gap: in GGA approximation the top of valence band is dominated by
Nickel $d$-states while in the GGA+U calculation the Oxygen $p$-states
give the most important contribution. In both approaches the bottom of
the conduction band is mainly Nickel $d$-like and therefore the predicted
band gap is primarily of charge-transfer type within GGA+U, in agreement
with experimental and theoretical evidence \cite{Lee91,wei1,Sawatzky84},
while it is wrongly described as of Mott-Hubbard type according GGA
approximation.

Our GGA+U value for the optical gap is $\approx$ 2.7 eV around the T
point, smaller that commonly accepted experimental values that range from
3.7 to 4.3 eV \cite{Powell70,Adler70,McNatt69,Hufner86}. More recently
however, a re-examination \cite{Hufner92} of the best available optical
absorption data \cite{Powell70} pointed out that optical absorption in NiO
starts at photon energy as low as 3.1 eV, not far from our theoretical
result. Indeed, Bengone and coworkers \cite{Bengone00} reported recently
an LDA+U calculation in NiO where different empirical values of $U$
were employed.  When $U=5$ eV was used---a value close to our present
first-principles result---, they obtained an optical gap of 2.8 eV,
very close to our results, {\it and} an excellent agreement between the
calculated and experimental \cite{Powell70} optical absorption spectra.
The same calculation with the literature value of $U=8$ eV, gave a
larger value for the optical gap but a very poor agreement with the
experimental absorption spectrum.

\subsection{Minerals: Fayalite}

As a final example we want to apply the present methodology to Fayalite,
the iron-rich end member of (Mg,Fe)$_2$SiO$_4$ olivine (orthorhombic
structure), one of the most abundant minerals in Earth's upper mantle.
Recently \cite{NoiFayalite} we showed that, although good structural
and magnetic properties could be obtained for this mineral within LDA or
GGA, its electronic properties were incorrectly described as metallic,
confirming the correlated origin of the observed insulating behavior.

\begin{figure}[t]
\includegraphics[width=9.0truecm]{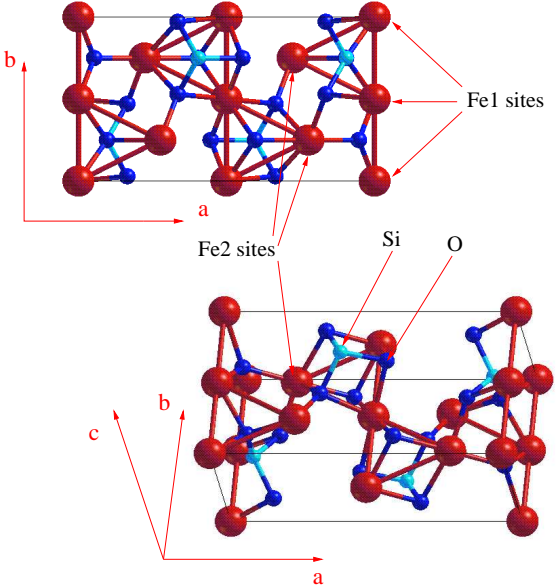}
\caption{\label{fig:str}The unit cell of Fayalite. Large dark ions are
Fe, small dark ions are O, light ions are Si.}
\end{figure}

From x-rays diffraction studies it is known that Fayalite has an
orthorhombic cell, whose experimental lattice parameters are (in atomic
units) a~=~19.79, b~=~11.50, c~=~9.11. The unit cell (depicted in fig.\
\ref{fig:str}) contains four formula units, 28 atoms: 8 iron, 4 silicon,
and 16 oxygen atoms. Silicon ions are tetrahedrally coordinated to
oxygens, whereas iron ions occupy the centers of distorted oxygen octahedra.
The point group symmetry of the non magnetic crystal is mmm (D$_{2h}$ in
the Schoenflies notation) and the space group is Pnma.  The magnetization
of iron reduces the original symmetry and only half of the symmetry
operations survive. The general expression for the internal structural
degrees of freedom is given in table \ref{tab:wyckoff} in the Wyckoff
notation \cite{Wyckoff}.

\begin{table}[h]
\caption{Definition of the Wyckoff structural parameters appropriate
for Fayalite structure}
\begin{tabular}{ccc}
 Ion & Class & Coordinates \\
 \hline
 Fe1 & 4a & (0,0,0), (1/2,0,1/2)\\
     &    & (0,1/2,0), (1/2,1/2,1/2) \\
 Fe2, Si, O1, O2 & 4c & $\pm$(u,1/4,v), \\
                  &    & $\pm$(u+1/2,1/4,1/2-v) \\
 O3 & 8d & $\pm$(x,y,z), $\pm$(x,1/2-y,z),\\
 & & $\pm$(x+1/2,1/2-y,1/2-z), \\
 & & $\pm$(x+1/2,y,1/2-z) \\
\end{tabular}
\label{tab:wyckoff}
\end{table}

Iron sites can be divided into two classes (see fig.\ \ref{fig:str}
and tab.\ \ref{tab:wyckoff}): Fe1 centers which are structured in chains
running parallel to the $b$, [010], side of the orthorhombic cell, and
Fe2 sites which belong to mirror planes for the non magnetic crystal
structure perpendicular to the $b$ side and cutting it at 1/4 and 3/4
of its length.  The main structural units are the iron centered oxygen
octahedra which are distorted from the cubic symmetry and tilted with
respect to each other both along the chains and on nearest Fe2 sites.
Fayalite is known to be an antiferromagnetic (AF) compound with slightly
non collinear arrangement of spin on Fe1 iron site (this non collinearity
will not be addressed here).  Magnetic moments along the central and the
edge Fe1 chains are antiferromagnetically oriented and  from our previous
work \cite{NoiFayalite} the most stable spin configuration is the one in
which the magnetization of Fe2 ion is parallel to the one of the closest
Fe1 iron. This magnetic structure is consistent with an iron-iron magnetic
interaction via a superexchange mechanism through oxygen $p$ orbitals.

\begin{figure}[!t]
\includegraphics[width=9.0truecm]{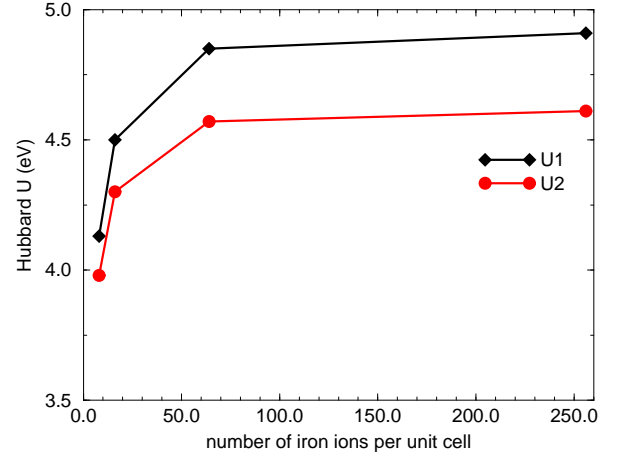}
\caption{\label{fayuconv}Convergence of Hubbard parameters of 
Fayalite with the number of iron included in the supercell used in the extrapolation. U1 is the value obtained for Fe1 ions, U2 the one for Fe2.}
\end{figure}

The calculation of $U$ was performed for the experimental geometry,
in the above mentioned spin configuration.  As the primitive unit
cell of fayalite is already quite large, we performed the constrained
calculation only in this cell and 
used larger supercells only to extrapolate the results. 
We considered three supercells in addition to
the primitive one: {\it i)} a cell duplicated in the $[0,1,0]$ chain
direction (a 1$\times$2$\times$1 supercell), containing 16 iron atoms;
{\it ii)} a cell, containing 64 iron ions, obtained by duplicating the
primitive structure in all directions (a 2$\times$2$\times$2 supercell)
and {\it iii)} a 4$\times$4$\times$2 supercell (256 iron ions).
Other computational details where similar to those used in our previous
work \cite{NoiFayalite}. As GGA approximation provided a slightly better
description of the system than LDA, we assumed this functional as the
starting point to be improved; the same pseudopotentials used in ref.\
\cite{NoiFayalite} for Fe, O and Si were adopted here; somehow larger
energy cutoff for the electronic wave functions and charge density
(36 and 288 Ry respectively) and a small smearing width of 0.005 Ry
were used. A 2$\times$4$\times$4 Monkhorst-Pack grid of k-points in the
primitive cell was found sufficient for the BZ integration.

The results of the $U$ calculation for the two different families of
iron sites (Fe1 and Fe2) are reported in fig.\ \ref{fayuconv} where
the rapid convergence with respect supercell dimension can be seen.
The final results for the on-site Coulomb parameters are $U_1=4.9$
eV for Fe1 ions and $U_2=4.6$ eV for Fe2, which are in fairly good
agreement with the approximate (average) value of 4.5 eV obtained in ref.\
\cite{NoiFayalite} from a rather crude estimate.

\begin{figure}[!t]
\includegraphics[width=10.0truecm]{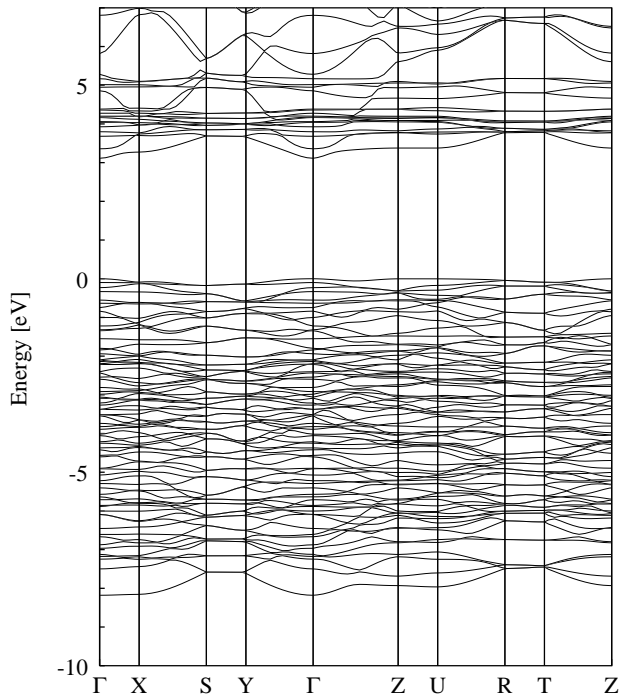}
\caption{\label{FayBANDS}
The band structure of Fayalite obtained within the present LDA+U approach.
The zero of the energy is set to the top of the valence band.
Complete degeneracy among spin up and spin down states is present.}
\end{figure}

\begin{figure}[!t]
\includegraphics[width=9.0truecm]{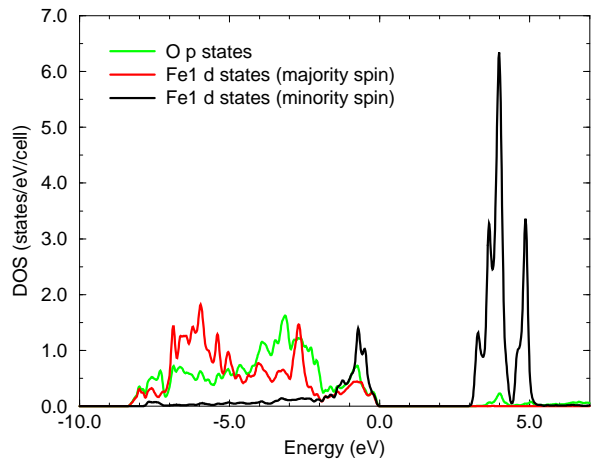}
\caption{\label{FayDOS}
Some atomic-projected density of states of Fayalite obtained within
the present LDA+U approach. Contributions from majority- and minority-spin 
$3d$ states of one of the Fe1 iron ions and from the total $2p$
manifold of one oxygen ion are shown.}
\end{figure}

The GGA+U band structure of Fayalite is shown in fig. \ref{FayBANDS}
while in fig.\ \ref{FayDOS} some atomic-projected density of states
are reported.  At variance with the GGA results reported in ref.\
\cite{NoiFayalite} a band gap of about 3 eV now separates the valence
manifold from the conduction one, in reasonable agreement with the
experimental result of about 2 eV \cite{fayexp1} at zero pressure.

The minority spin $t_{2g}$ manifold of iron ions, that within GGA
cross the Fermi energy, is split into two subgroups by the gap opening.
The conduction-band states are shrunk to a narrow energy range and
moved above the bottom of the iron $s$-states band which remains almost
unaffected; the lower-energy minority-spin $d$-states, instead, merge
in the group of states below the Fermi level where they mix strongly
with states originating from Oxygen $p$ orbitals: the two sets of
states, well separated in the GGA results, collapse into a unique block.
The most evident consequence of the gap opening consists in a pronounced
shrinking of the $d$ states of iron which become flatter than in the
GGA case. This is evident on the top of the valence band, but also for
states well below this energy level, which thus reveal a more pronounced
atomic-like behavior.  Beside the gap opening between the two groups
of the minority-spin states, a strong mixing occurs among the oxygen $p$
states and the iron $d$ levels over a rather large region extending down
to 8 eV below the top of the valence band. A finite contribution of
the oxygen states is present close to the top of the valence manifold
showing that the gap is mainly of Mott-Hubbard type with a partial
charge-transfer character.

We have then relaxed the geometric structure of the system (both internal
and cell degrees of freedom) assuming no dependence of $U_1$ and $U_2$ on
the atomic configuration. The resulting structural parameters ($a=20.18$,
$b=11.75$, $c=9.29$ atomic unit) as well as the internal coordinates
reported in table \ref{tab:faystrut} are in very good agreement with the
experimental results, even better than the already satisfactory agreement
obtained in ref.\ \cite{NoiFayalite} within GGA.

\begin{table}[t]
\caption{Comparison of the experimental and LDA+U calculated values
for the Wyckoff structural parameters of Fayalite as defined in table
\protect\ref{tab:wyckoff}}
\begin{tabular}{cc|ccccc}
& Ion & u & v & x & y & z \\
\hline
Exp.\ & & \\
 &Fe2 &0.780 &0.515 & & &  \\
 &Si  &0.598 &0.071 & & &  \\
 &O1  &0.593 &0.731 & & &  \\
 &O2  &0.953 &0.292 & & &  \\
 &O3  & & &0.164 &0.038 & 0.289 \\
$GGA+U$ & & \\
 &Fe2 &0.779 &0.515 & & &  \\
 &Si  &0.597 &0.072 & & &  \\
 &O1  &0.593 &0.735 & & &  \\
 &O2  &0.951 &0.289 & & &  \\
 &O3  & & &0.165 &0.036 & 0.286 \\
\end{tabular}
\label{tab:faystrut}
\end{table}

Although we did not studied other spin-configurations, magnetic
properties seam to improve slightly in the GGA+U approximation.
The magnetic moment on each iron (both Fe1 and Fe2) was found to be 3.9
$\mu_{B}$, in closer agreement with the spin-only value (4 $\mu_{B}$)
of the experimental result (4.4 $\mu_{B}$) than the one obtained by GGA
only (3.8 $\mu_{B}$). This improvement is probably due to the enhanced
atomic-like character of iron $d$ states, which is consequence of the
gap opening.

In conclusion, the GGA+U provides a quite good description of structural,
magnetic {\it and} electronic properties of fayalite, reproducing the
observed insulating behavior with a reasonable value for its fundamental
band gap.

\section{Summary}

In this work we have reexamined the LDA+U approximation to DFT and a
simplified rotational-invariant form of the functional was adopted. We
then developed a method, based on a linear response approach, to calculate
in an internally consistent way the interaction parameters entering the
LDA+U functional, without making aprioristic assumption about screening
and/or basis set employed in the calculation. Our methodology was
then successfully tested on a few systems representative of normal and
correlated metals, simple transition metal oxides and iron silicates.
In all cases we obtained rather accurate results indicating that our
scheme allows us to study both electronic and structural properties of
strongly correlated material on equal footing, without resorting to
any empirical parameter adjustment.

\begin{acknowledgments}
This work has been supported by the MIUR under the PRIN program and
by the INFM in the framework of the {\it Iniziativa Trasversale Calcolo
Parallelo}.
\end{acknowledgments}


\vfill\eject

\end{document}